# Widely FSR tunable high Q-factor microresonators formed at the intersection of straight optical fibers

**ISHA SHARMA[1],*  AND MISHA SUMETSKY[1],†**

[1]*Aston Institute of Photonic Technologies, Aston University, Birmingham B4 7ET, UK*
*i.sharma1@aston.ac.uk, †m.sumetsky@aston.ac.uk

**We present a new class of high-Q tunable microresonators formed at the intersection of two straight silica optical fibers, whose free spectral range (FSR) can be widely tuned by fiber rotation. The proposed configuration avoids the limitations of traditional monolithic microresonators that lack FSR tunability required for a wide range of photonic applications. Using small rotation angles (1–15 mrad), we demonstrate a tunability of the FSR from 2 pm to 10 pm, enabled by microscale fiber displacements that reshape the resonator profile over millimeter scales. The proposed approach minimizes mechanical stress, supports miniaturization, and is suitable for integration with MEMS. It paves the way for the fabrication of tunable delay lines, ultralow repetition rate broadband frequency comb generators, and nonlocal optofluidic sensors on a chip.**

Optical microresonators are essential components in various types of miniature photonic devices due to their small dimensions, high Q-factor, and unique spectral characteristics [1-3]. However, the solid monolithic structure of commonly used optical microresonators limits their tunability, which is critical for applications including cavity quantum electrodynamics [4, 5], optical classical and quantum signal processing and computing [6, 7], and frequency comb generation [8, 9]. Existing tuning approaches, like those based on mechanical, thermal, electro-optical, and nonlinear optical effects [10-14], typically shift the microresonator eigenwavelengths jointly, without significant change of their free spectral range (FSR). Tuning the FSR of microresonators, being of particular importance for a wide range of applications [4-9], is especially challenging since it requires substantial modification of resonator dimensions or refractive index, which is difficult to achieve in compact monolithic designs.

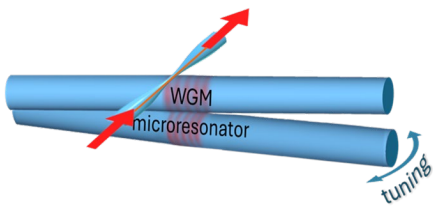

**Fig. 1.** A WGM microresonator induced at the silica fiber intersection and excited by a microfiber taper.

Alternatively, Surface Nanoscale Axial Photonics (SNAP) platform, where whispering gallery mode (WGM) microresonators are formed by nanoscale deformation of optical fibers [15], enables the creation of reconfigurable resonators with tunable eigenwavelength separations [16-18]. One method to tune the FSR of a SNAP microresonator is to employ the local stress-induced refractive index variation by bending an optical fiber [16, 17]. Yet, the stress vanishes with the growing of the bending radius, leading to the disappearance of longer resonators. Alternatively, several millimeters long FSR tunable high Q-factor SNAP resonators, important for a range of applications discussed below, can be formed by coupling co-planar bent fibers [18]. Miniaturization of design [18] targeting on-chip applications can be realized through the micron-scale MEMS manipulation of fiber ends (see Fig. 1 in Ref. [18]). However, the required configuration of coupled fibers and an input-output waveguide, which enable millimeter-scale tunability of the coupling length, is challenging to realize on a chip with minimal applied forces.

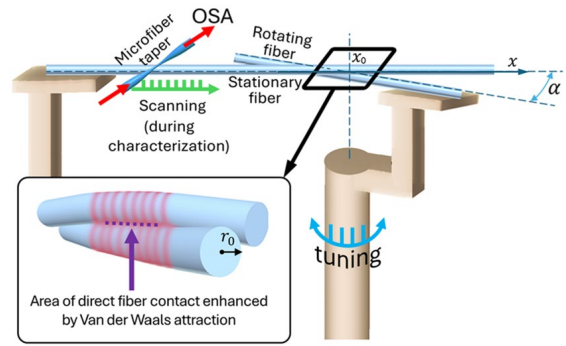

**Fig. 2.** Illustration of the experimental setup.

To address this challenge, here we propose a new microresonator geometry: a pair of straight silica fibers intersecting at a tunable angle illustrated in Fig. 1. Through controlled rotation, this simple and compact system, which can be readily transferred to the microscale with minimal actuation forces, enables a wide range of FSR tuning. Specifically, we experimentally demonstrate that *dramatically small rotation angles ≲ 10 mrad, which*

*correspond to ≲ 10 µm fiber shifts of a few mm long fiber, enable millimeter-scale variations in the resonator length and an order of magnitude picometer-scale FSR tuning.*

In our experiment, we used two silica fiber segments with radius $r_0 = 62.5$ µm and refractive index $n_0 = 1.44$. After the fiber segments were uncoated and cleaned in methanol and flame, they were arranged to intersect at a rotation angle $\alpha$, which was tuned by rotating one of the fibers as illustrated in Fig. 2. The spectra of this configuration were measured along the fiber length using a tapered fiber with a micron-scale waist. The tapered fiber, connected to a BOSA 400 optical spectrum analyzer, made periodic contacts with the stationary fiber at intervals of 20 µm along its axis (see Fig. 2). The measured spectra were ordered in 2D spectrograms, presenting the normalized transmission power as a function of the longitudinal coordinate $x$ along the stationary fiber and wavelength $\lambda$ (see e.g., [15]). Exemplary spectrograms are shown in Fig. 3 for various intersection angles $\alpha$ in the vicinity of a cutoff wavelength (CWL). The CWL (a red trail tracking the dips in the transmission spectrum) was chosen near $\lambda = 1556$ nm and corresponds to the WGM azimuthal quantum number $m \cong 2\pi r_0 n_0/\lambda \cong 360$. This particular CWL was selected as a representative from a series of CWLs corresponding to various WGM azimuthal and radial quantum numbers (see Supplementary Material).

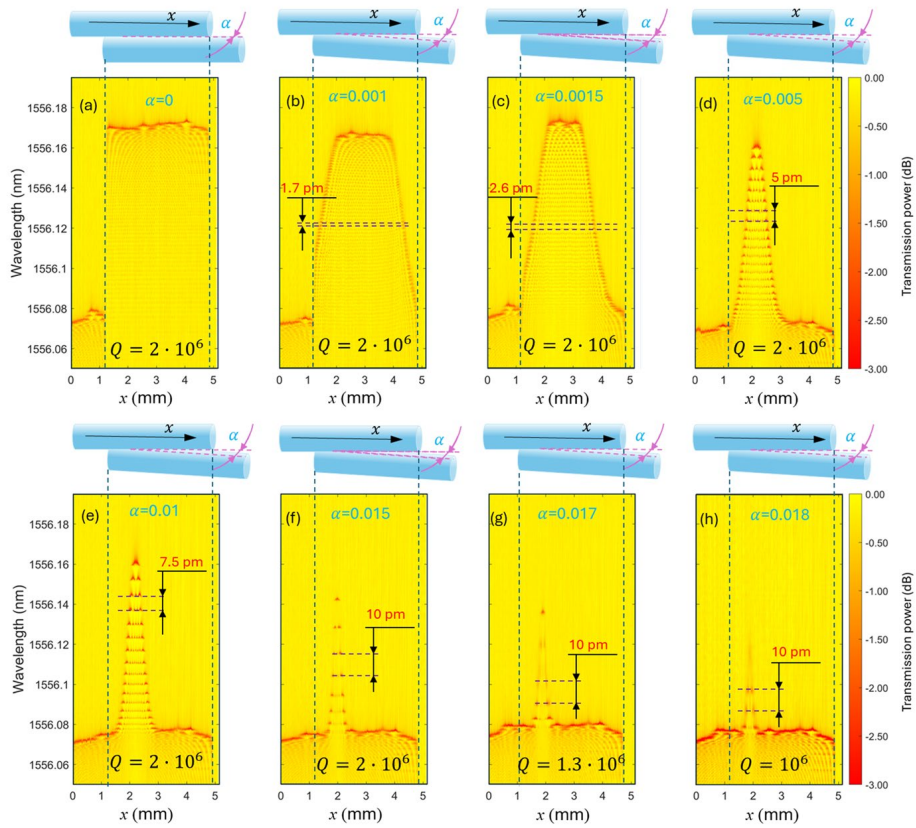

**Fig. 3.** (a)-(h) spectrograms of the induced SNAP microresonators for different intersection angles $\alpha$.

Fig. 3(a) shows the spectrogram for $\alpha = 0$, i.e., for the parallel fibers. The fibers are side coupled along their 3.6 mm length, at $1.2 < x < 4.8$ mm. The region $0 < x < 1.2$ mm in this spectrogram corresponds to the CWL variation along the fiber axis, $\lambda_{CWL}(x)$, (bold red colored) for the uncoupled upper (stationary) fiber. Minor noise in $\lambda_{CWL}(x)$ is due to the uncontrolled environment in our lab causing fiber surface contamination. Fiber coupling in the region $1.2 < x < 4.8$ mm introduces the shift of $\lambda_{CWL}(x)$ by ~ 0.1 nm. This spectrogram is a part of the spectrogram along a broader wavelength spectrum presented in Fig. S1 of the Supplementary Material, where the CWLs of the upper fiber ($x < 1.2$ mm), coupled fibers ($1.2 < x < 4.8$ mm) as well as lower (rotating) fiber ($x > 4.8$ mm) are visible.

We observed the appearance of a high Q-factor microresonator at very small intersection angles $\alpha \sim 1$ mrad presented by the spectrograms in Fig 3(b) and (c) for $\alpha = 1$ mrad and $\alpha = 1.5$ mrad, respectively. Owing to a relatively long coupling region, these microresonators have small FSRs ~ 1-3 pm. Due to the uncontrolled environment in our lab, the loaded Q-factor in both cases was measured as $Q \cong 2 \cdot 10^6$, which we suggest can be improved to $Q \sim 10^8$ in a clean room

[12, 19]. In Fig. 3(b), the FSR slowly grows with distance from the top of the $\lambda_{CWL}(x)$, since the resonator CWL profile is close to rectangular. Alternatively, in Fig. 3(c) the FSR is close to constant for a large portion of eigenwavelengths. This observation is confirmed by a model based on the SNAP microresonator theory [15]. According to this theory, the eigenwavelengths and eigenmodes of the induced microresonator are described by the Schrödinger equation, where the CWL variation serves as a potential. In the comparison of our experimental data with theory shown in Fig. 4, we modeled the CWL variation by a profile having a flat top surrounded by parabolic sides. For clearer visualization, we have added a uniformly spaced template covering both the experimental spectrogram and the model spectrum in this figure. We found that the FSR of our model is accurately uniform in a significant middle part and deviates from uniformity in the vicinity of the top and bottom of the resonator's spectrum. Notably, while the potentials having a *globally* constant FSR (like, e.g., a parabolic potential) have been comprehensively investigated [20-22], to our knowledge, a class of potentials exhibiting a constant FSR over *a portion* of their spectrum, important for applications, has not been studied previously. This interesting problem will be considered elsewhere [23].

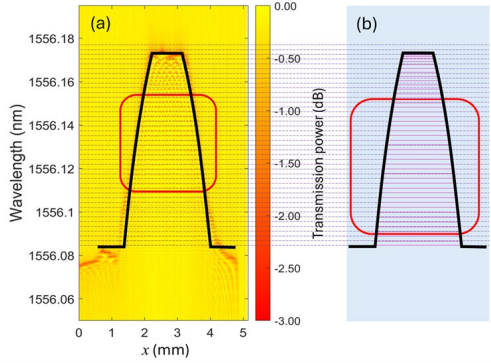

**Fig. 4.** (a) The spectrogram of microresonator reproduced from the spectrogram in Fig. 3(c). (b) The spectrum of a model resonator with flat top and parabolic sides. The FSRs in Figs. (a) and (b) are compared using a background template of horizontal equidistant parallel dashed lines.

Remarkably, the measured CWL profile shown in Fig. 3(b) does not monotonically grow with the separation between fibers. Instead, the CWL behavior indicates that the fibers are in direct contact along an ~ 2 mm long interval ($2 < x < 4$ mm) and start to separate beyond this interval only. It is seen from the spectrogram shown in Fig. 3(c) for $\alpha = 1.5$ mrad that the length of direct contact reduces with increasing $\alpha$. We suggest that in these intervals the fibers stay in direct contact as illustrated in the inset of Fig. 2 due to their strong attraction. Indeed, in the absence of the interaction between fibers, they remain straight and their separation $d(x)$ as a function of the distance from the intersection point $x = x_0$ (Fig. 2) is calculated for $\alpha \ll 1$ as

$$d(x) = \frac{\alpha^2(x-x_0)^2}{4r_0}. \quad (1)$$

This dependence is shown in Fig. 5 for different intersection angles $\alpha$. It is seen from Fig. 5 that, for $\alpha = 1$ mrad, the separation between fibers is less or comparable with 1 angstrom over the distance of ~ 0.5 mm. The attraction between fibers at such dramatically small separation becomes very strong (primarily dominated by Van der Waals forces [24, 25]) and causes the fibers to come into direct contact over a length scale on the order of a millimeter, as confirmed by the spectrograms in Figs. 3(b) and (c). We suggest that the observed fiber attraction offers a unique optical approach to study the interaction between silica surfaces at angstrom-scale separations.

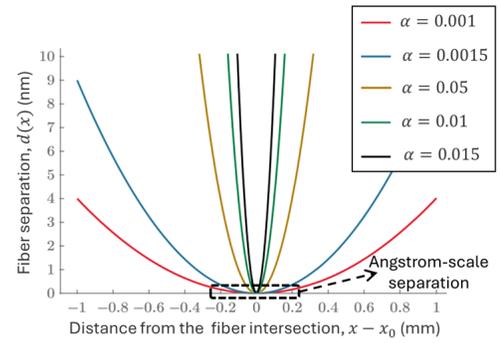

**Fig. 5.** Fiber separation, $d(x)$, as a function of distance from the fiber intersection, $x - x_0$, for different intersection angles $\alpha$.

For larger intersection angles $\alpha = 5, 10,$ and 15 mrad, the microresonator spectrograms are shown in Figs. 3(d), (e), and (f). It is seen that the length of microresonators gradually decreases with increasing $\alpha$, while the effect of the attraction vanishes. Correspondingly, the axial FSR of microresonators increases from ~ 5 pm for $\alpha = 5$ mrad to ~ 10 pm for $\alpha = 15$ mrad. In contrast to the behavior of FSR in Fig. 3(b), the FSRs in the spectrograms of Figs. 3(d), (e), and (f) decrease with the distance from the top of the $\lambda_{CWL}(x)$ due to the qualitatively different CWL profiles. The Q-factors of these microresonators remain the same, $Q \cong 2 \cdot 10^6$.

At the intersection angles $\alpha \cong 15$ mrad, we observed degradation of the induced microresonators spectra accompanied with the reduction of their Q-factors. This is presumably caused by a reduction of light getting back to the resonator after circulation in the mutually angled fibers. The exemplary behavior of the resonator modes is presented by the spectrograms in Figs. 3(g) and (h). Remarkably, the behavior of modes in these spectrograms does not clarify the tendency of their disappearance. For example, in Fig. 3(g) the fundamental axial mode remains strong together with mode 5, while modes 2, 3, and 4 are noticeably weakened. This suggests a strong effect of interference, rather than mere attenuation of the recirculating power, which can be explained by the prospective theoretical study and numerical modelling of the proposed fiber configuration.

The spectrograms shown in Fig. 3 demonstrate that the picometer-scale FSR of our microresonator can be widely tuned by micron-scale displacement. In particular, for the fiber lengths of our concern $L \sim 1$ mm and rotation angles $\alpha \lesssim 10$ mrad, the required fiber end displacement is $\alpha L \lesssim 10$ μm. Given the feasibility of achieving such small displacements with minimal actuation force using MEMS [26], along with the development of methods for coupling fiber-supported WGMs to on-chip waveguides [27-29], we suggest that the demonstrated tunable microresonators can be integrated and precisely controlled on a chip.

The introduced class of tunable microresonators enables a range of *tunable photonic devices integrated on a chip* that, to our knowledge, were previously unattainable. These devices include tunable delay lines, broadband tunable low repetition rate optical frequency comb (OFC) generators, and tunable nonlocal optofluidic sensors.

In particular, similar to previously demonstrated *stationary* designs of slow WGM delay lines on a single fiber [30, 31], a *tunable delay line* based on the fiber configuration considered here can be created. A WGM light pulse slowly propagating along the axis of the stationary fiber (Fig. 2) reflects at the CWL turning point and returns to the input-output waveguide after the nanosecond-scale delay time. The position of the turning point and, consequently, the delay time is controlled by the angle of rotation. Critically, a bandwidth with a close to uniform FSR (like that in the spectrogram in Fig. 3(d)), which ensures the dispersionless pulse propagation [30], will remain locally constant for wider resonators with smaller FSR and longer time delays [31] without shrinking the bandwidth as in the case of delay line based on single fiber bending [16, 17, 32].

Another feasible application is a *broadband tunable low repetition rate OFC generator*. This device can be created by connecting our microresonator, which has a tunable picometer-scale FSR $\Delta\lambda_t$, with another microresonator possessing a much higher broadband FSR $\Delta\lambda_h \gg \Delta\lambda_t$. An analogous stationary configuration of a broadband OFC generator has been proposed in Ref. [33] for a bottle microresonator and experimentally demonstrated recently based on the thin film lithium niobate photonic platform in Ref. [34]. Under the matching condition $\Delta\lambda_h = N\Delta\lambda_t$ with integer $N \gg 1$, this device can generate a broadband, low-repetition-rate frequency comb [33, 34] that can be finely tuned by varying $\Delta\lambda_t$ and $N$ while preserving $N\Delta\lambda_t$. Of particular interest is the exploration of a similar configuration that employs highly nonlinear low-loss fibers, such as lithium niobate fibers polished to optical-grade precision.

Replacing one of the fibers with a microcapillary transforms the proposed configuration into a *tunable nonlocal microfluidic sensor*, operating by tracking shifts of the resonant eigenwavelengths. Beyond the capabilities of similar stationary nonlocal microfluidic sensors [35, 36], this device enables the characterization of more complex microfluidic phenomena by collecting enriched data from microresonators with tunable shapes.

More complex tunable CWL profiles leading to a different FSR behavior can be fabricated by introducing permanent bends and nanoscale effective fiber radius variations [15, 30] near the intersection of fibers with similar or different radii. We also suggest that moving from the uncontrolled environment of our lab to a clean environment will allow us to increase the measured loaded Q-factor $Q \cong 2 \cdot 10^6$ to $Q \sim 10^8$ [12, 19]. Finally, the development of the theory and numerical modeling of the proposed new geometry of a WGM microresonator is of great interest.

**Funding.** Engineering and Physical Sciences Research Council (EPSRC) grants EP/W002868/1 and EP/X03772X/1, Leverhulme Trust grant RPG-2022-014.

**Disclosures.** The authors declare no conflicts of interest.

**Data availability.** Data underlying the results presented in this paper may be obtained from the corresponding author upon reasonable request.

**Supplemental document.** See Supplemental Material for supporting content.

# Supplemental Material

## Spectrograms of coupled fibers exhibiting multiple CWLs

Fig. S1(a) shows a spectrogram of two parallel fibers corresponding to intersection angle $\alpha = 0$ along the wavelength bandwidth from 1553.5 nm to 1558.5 nm. The spectrogram is normalized to the transmission power value outside of the resonance region. The fibers are side coupled along their 3.6 mm length, for $1.2 < x < 4.8$ mm. The region $x < 1.2$ mm corresponds to the uncoupled upper (stationary) fiber, while the region $x > 4.8$ mm corresponds to the lower (rotating) fiber. Series of CWLs, different for uncoupled fibers (regions $x < 1.2$ mm and $x > 4.8$ mm) and coupled fibers (region $1.2 < x < 4.8$ mm) are indicated by red colored trails of dips in the transmission spectrum. A section of this spectrogram including the CWL situated close to 1556.1 nm, outlined by a rectangle, is selected and magnified in the main text to monitor the behavior of the microresonator spectra for different intersection angles $\alpha$. Accordingly, a section outlined by the rectangle in Fig. S1(a) is magnified in Fig. 3(a). Similarly, for $\alpha = 5$ mrad, a section of the spectrogram shown in Fig. S1(b) outlined by the black rectangle is shown in Fig. 3(d).

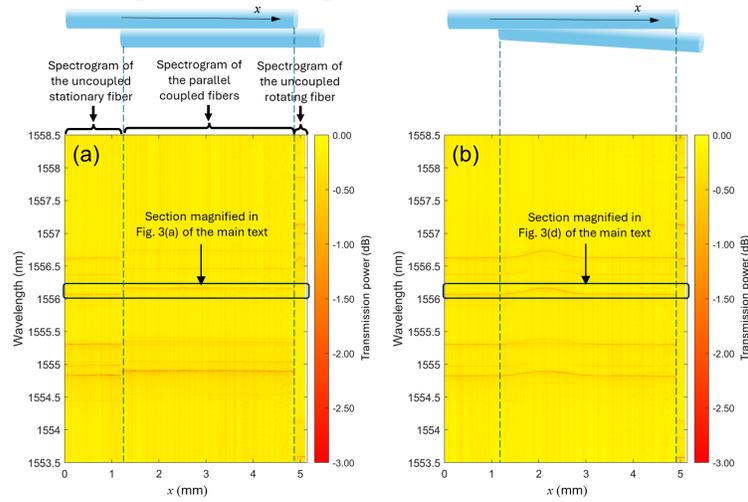

**Fig. S1.** (a) A spectrogram of parallel fibers coupled along the 3.6 mm of their length. (b) A spectrogram of fibers intersecting under the angle $\alpha = 5$ mrad.

## Correction of temperature-dependent deviations

The spectrograms were measured along the 5.1 mm length of the fiber with 20 μm resolution over the time of 64 min, corresponding to 15 sec of the single measurement time. The temperature of the environment and fibers slowly changed in the process of measurements. This change affected the measured spectra. Since the wavelengths of the spectral resonances should remain constant for constant temperature, we attribute the spectral shifts to the temperature variations, which can be simply corrected. As an example, Fig. S2(a) shows the originally measured spectrogram of our fiber configuration for $\alpha = 5$ mrad. The magnified section of this spectrogram in Fig. S2(a1) shows the spectral shifts growing with time. From Fig. S2(a1) we find that the change of the resonance positions is close to linear in time and is $\delta\lambda \sim 1.3$ pm over the scanning distance of 1.6 mm, which was completed in time $\tau \sim 19$ min. The temperature variation during this time is found as $\left(\frac{\delta\lambda}{\lambda}\right)\left(\frac{dn}{dT}\right)^{-1} \sim 0.08$K, where we used $\frac{dn}{dT} \cong 10^{-5}$K$^{-1}$ for silica. To correct for this temperature dependence, we linearly shift the measured spectra vertically to arrive at the spectrogram shown in Figs. S2(b) and (b1). We performed similar linear shifts correcting the temperature variations on the other spectrograms of Fig. 3.

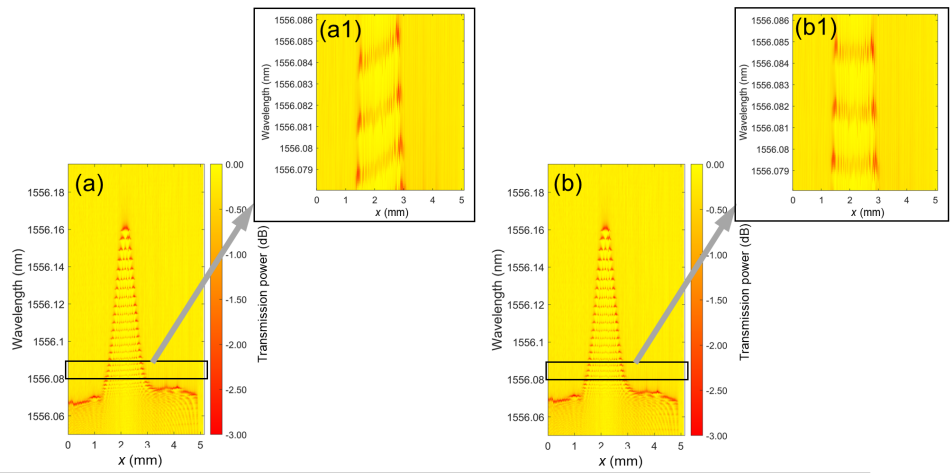

**Fig. S2.** (a) Originally scanned spectrogram of a microresonator corresponding to the intersection angle $\alpha = 5$ mrad with magnified section of this spectrogram shown in (a1). (b) Spectrogram shown in (a) corrected for the temperature variation with the magnified section shown in (b1).